\documentclass[prl,twocolumn]{revtex4}
\pdfoutput=1
\usepackage{amsfonts,amssymb,amsmath}            
\usepackage[]{graphics,graphicx,epsfig}            
\usepackage{amsthm}
\def\identity{\leavevmode\hbox{\small1\kern-3.8pt\normalsize1}}
\newcommand{\ket}[1]{\left | #1 \right\rangle}
\newcommand{\bra}[1]{\left \langle #1 \right |}
\newcommand{\half}{\mbox{$\textstyle \frac{1}{2}$}}
\newcommand{\smallfrac}[2][1]{\mbox{$\textstyle \frac{#1}{#2}$}}

\newcommand{\proj}[1]{\ket{#1}\bra{#1}}

\renewcommand{\epsilon}{\varepsilon}
\begin{document}

\title{Perfect Quantum Routing in Regular Spin Networks}
\date{\today}

\author{Peter J.~\surname{Pemberton-Ross}}
\affiliation{Centre for Quantum Computation,
             DAMTP,
             Centre for Mathematical Sciences,
             University of Cambridge,
             Wilberforce Road,
             Cambridge CB3 0WA, UK}
\author{Alastair \surname{Kay}}
\affiliation{Centre for Quantum Computation,
             DAMTP,
             Centre for Mathematical Sciences,
             University of Cambridge,
             Wilberforce Road,
             Cambridge CB3 0WA, UK}
             
\begin{abstract}
Regular families of coupled quantum networks are described such the unknown state of a qubit can be perfectly routed from any node to any other node in a time linear in the distance. Unlike previous constructions, the transfer can be achieved perfectly on a network that is local on any specified number of spatial dimensions. The ability to route the state, and the regularity of the networks, vastly improve the utility of this scheme in comparison to perfect state transfer schemes. The structures can also be used for entanglement generation.
\end{abstract}

\maketitle

{\em Introduction:} The task of quantum state transfer was introduced in the context of quantum computation as a protocol to simplify interactions between distant qubits in an architecture that has locality restrictions, as in solid state systems. This study was initiated by Bose who analysed a uniformly coupled quantum chain, and evaluated its efficacy for transferring an unknown quantum state from one end to the other \cite{Bos03}. A plethora of protocols have since been introduced to achieve transfer perfectly \cite{Christandl,lambrop,transfer_comment,Kay:2005e} or with arbitrary accuracy \cite{sougato_review}.

The only perfect transfer protocols that do not directly couple every qubit in the network \cite{niko}, and hence do not have trivial transfer distance, were designed to transfer a quantum state which is input on a given site, onto a specific, corresponding output site. These schemes are either highly non-local (hypercubes \cite{Kay:2004c,facer,simone2} and integral circulant graphs \cite{simone3}) or local, but irregular. Both of these factors severely reduce the interest in implementing state transfer protocols in the lab, and the usefulness of state transfer has become more apparent as a constructive tool \cite{kay:review,Kay:09}. Both of these problems have been overcome in the arbitrarily accurate scenario by reintroducing some control on the input and output spins \cite{wojcik,johan,hanggi}. In fact, this reintroduction is not unreasonable, since, although it is desirable to assume no interaction with the system, and just let its (fixed) Hamiltonian generate the transfer, this is to forget one vital element of a state transfer protocol -- it is assumed that one can introduce the quantum state onto the input node, and remove it from the output node, which needs to be implemented quickly in comparison to the Hamiltonian dynamics (although this `quickly' restriction can also be reduced \cite{haselgrove04,Kay:08b}). The vastly richer dynamics due to this additional control was recently noted in \cite{Kay:10}.

In this paper, we make the same assumption -- that local gates can be applied quickly on the output nodes, and use this to design protocols to transfer a quantum state from any node to any other node in a regular network of arbitrary spatial dimension. We refer to this task as {\em perfect routing}. In doing so, we overcome some of the major limitations of previous perfect state transfer schemes, and introduce some robustness to manufacturing imperfections. There are also several advantages over the previous arbitrarily accurate schemes in terms of scaling properties of the required fields, and the ability to route multiple states at once. The motivation is similar to the cluster state methodology, allowing local gates to manipulate a pre-specified resource, in this case a propagating Hamiltonian. In comparison to using an Ising interaction to generate a cluster state, which can be used to generate a maximally entangled pair between any pair of nodes, so that a state can then be teleported, our scheme will not require any feed-forward of measurement results, allowing the required control to be periodic in both space and time.

\begin{figure}[!bt]
\begin{center}
\includegraphics[width=0.4\textwidth]{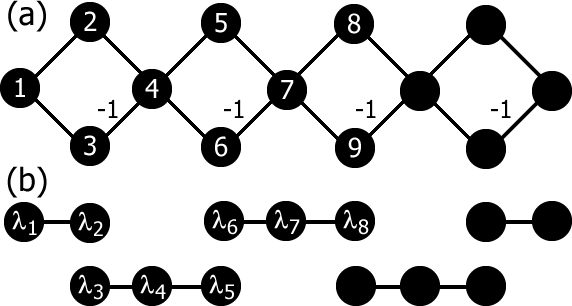}
\end{center}
\vspace{-0.5cm}
\caption{(a) A quasi-1D routing structure. The circles represent qubits, and the lines indicate an $XX$ coupling between pairs of qubits of strength $+1$, unless $-1$ is indicated. (b) Under a basis transformation, a simple direct sum structure is apparent. In this case, all coupling strengths are $\sqrt{2}$.} \label{fig:1D}\vspace{-0.5cm}
\end{figure}

{\em One-dimensional prototype:} We start by considering routing in a one-dimensional system of $3N+1$ qubits, as depicted in Fig.~\ref{fig:1D}(a), in order to illustrate some of the basic ideas of our construction. The fixed Hamiltonian takes the form
$$
H=\half\sum_{\{n,m\}\in E}J_{n,m}(X_nX_m+Y_nY_n)
$$
where $E$ is the set of edges of the graph depicted in Fig.~\ref{fig:1D}(a) and $J_{n,m}=1$, unless $\{n,m\}=\{3k,3k+1\}$ for some $k$, in which case $J=-1$ ($\hbar=1$ such that times and energies can be treated as dimensionless). The Hamiltonian is spin preserving,
$$
\left[H,\sum_{n=1}^{3N+1}Z_n\right]=0,
$$
so we analyse a protocol where all spins are initialised in the $\ket{0}$ state. By introducing the state to be transferred on a particular spin, we place the system in a superposition of 0 and 1 excitations. The 0 excitation subspace is a single state, which is therefore invariant under the Hamiltonian evolution, and thus we can concentrate on the single excitation subspace. We denote the basis states by $\ket{n}=\ket{0}^{\otimes n-1}\ket{1}\ket{0}^{\otimes 3N+1-n}$. The crucial property that we make use of here is that, for example, $(\ket{2}+\ket{3})/\sqrt{2}$ is a 0 eigenstate of $(XX+YY)_{2,4}-(XX+YY)_{3,4}$, and similarly $(\ket{2}-\ket{3})/\sqrt{2}$ is a 0 eigenstate of $(XX+YY)_{1,2}+(XX+YY)_{1,3}$. This means that we can rewrite the basis states as $\ket{\lambda_{3n+1}}=\ket{3n+1}$, $\ket{\lambda_{3n+2}}=(\ket{3n+2}+\ket{3n+3})/\sqrt{2}$ and $\ket{\lambda_{3n+3}}=(\ket{3n+2}-\ket{3n+3})/\sqrt{2}$, leaving the Hamiltonian with a direct sum structure as depicted in Fig.~\ref{fig:1D}(b). Each subsystem is a uniformly coupled chain of length 2 or 3, and achieves perfect transfer in times $\pi/\sqrt{2}$ and $\pi/2$ respectively. So, starting with a state $\ket{\lambda_{3n}}$, after a time $\pi/2$, we have $\ket{\lambda_{3n+2}}$. Now, observe that a fast application of the local rotations (but globally applied) $U=\prod_{n=1}^NZ_{3n}$ converts between states $\ket{\lambda_{3n+2}}$ and $\ket{\lambda_{3n+3}}$, i.e.~it transfers the state from one subsystem to the next. Hence, starting from $\ket{\lambda_{3n+2}}$, we apply $U$ every $\pi/2$ and after $|m-n|\pi/2$, we arrive in the state $\ket{3m+2}$. This achieves the long range transfer, and we just have to show, at the start, how to convert from the input state, either $\ket{3n+1}$ or $\ket{3n+2}$, to $\ket{\lambda_{3n+2}}$. This step also has to be inverted in the end, but the periodic dynamics ensure this.

If the input qubit is 1, then $\ket{\lambda_2}$ is simply produced by letting the Hamiltonian evolve for time $\pi/\sqrt{2}$. This operation is its own inverse. For other starting qubits $3n+1$, we apply the evolution $e^{-iH\pi/4}Z_{3n+2}Z_{3n+3}e^{-i3H\pi/8}$. (Note that $Z_{3n+2}Z_{3n+3}$ is equivalent to applying a $Z$ rotation to the $3n+2$ end of the 3-qubit effective chain in Fig.~\ref{fig:1D}(b).) More crucially for the coming generalisation, we must show how to start from the vertices of the diamonds, say $\ket{3n+2}$, and produce $\ket{\lambda_{3n+2}}$. Provided $n\neq 0,N-1$, we can simply evolve the Hamiltonian for a time $\pi/2$, at which point we have
$$
(\ket{\lambda_{3n}}+\ket{\lambda_{3n+5}})/\sqrt{2}.
$$
Upon application of phase gates $\sqrt{Z_{3n-1}Z_{3n}}$, we are left with
$$
(i\ket{\lambda_{3n}}+\ket{\lambda_{3n+5}})/\sqrt{2}.
$$
Subsequent evolution for a time $\pi/2$ converts the state to
$$
(i\ket{\lambda_{3n+2}}+\ket{\lambda_{3n+3}})/\sqrt{2}=(\ket{3n+2}-i\ket{3n+3})/\sqrt{2},
$$
and a phase gate $\sqrt{Z_{3n+3}}$ yields $\ket{\lambda_{3n+2}}$.

This suffices to prove that, with the help of local magnetic fields, we can perfectly propagate an unknown quantum state along the length of the chain, between any set of nodes in a time proportional to the distance between the nodes. Since the subsystems in Fig.~\ref{fig:1D}(b) are independent, we can actually have multiple excitations in the system provided they are each separated by at least one subsystem. Within a 1D structure, this imposes that one state can never move past a second one. To resolve this, we need to move to a two or three dimensional structure, as we will now demonstrate.

\begin{figure}[!tb]
\begin{center}
\includegraphics[width=0.45\textwidth]{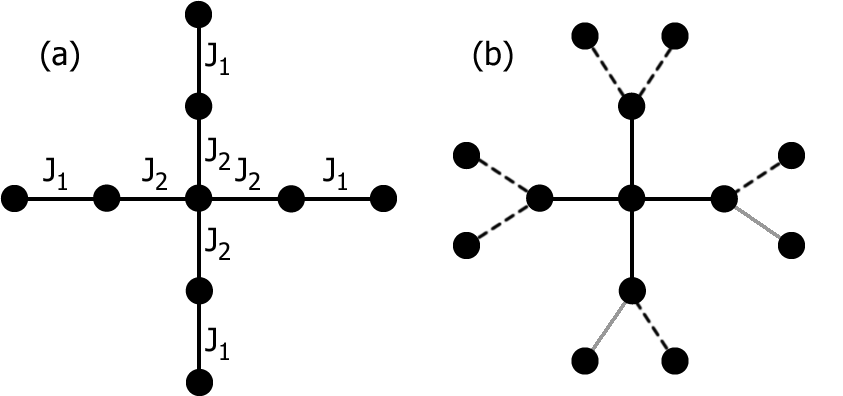}
\end{center}\vspace{-0.5cm}
\caption{The basic scheme for a 2D square lattice. Once perfect routing between extremal nodes in (a) has been shown, this can be converted into a repeating unit (b) which can be tiled to give a complete network which has a direct sum structure given by (a). Dashed edges denote a scaling factor of coupling strengths by $1/\sqrt{2}$ relative to (a). The gray lines are scaled by $-1/\sqrt{2}$.} \label{fig:2D}\vspace{-0.5cm}
\end{figure}

{\em Designing Systems with a Subsystem Reduction:} Figure \ref{fig:2D} demonstrates the straightforward generalisation of the 1D results so that we can make a full network out of any basic building block that we want to be the structure of a subsystem. We simply take the extremal links in a network, which are coupled with strength $J$, and replace them with a $V$ structure such that each coupling strength has modulus $J/\sqrt{2}$. These couplings are then patterned such that when the structure is tiled, one of the 4 coupling strengths in each diamond is $-J/\sqrt{2}$. This is depicted for 2D, but works for any dimension $d$. As before, applying a $Z$ rotation on one of the extremal spins of a block hops an excitation present across the V between $(\ket{01}+\ket{10})$ and $(\ket{01}-\ket{10})$, which are effective single excitations on two independent subsystems. Thus, we are simply tasked with demonstrating how to route from any extremal node of the subsystem to any other. Similarly to the 1D case, multiple states can be transferred at once, since all states are kept confined to individual sections; it suffices to keep these sections separated by a single unit, and then the states never meet, never interfering.

The subsystem structure also allows a simple consideration of the effect of some errors. For instance, if manufacturing errors can be identified on a given subsystem, then provided these errors are not on the input or output blocks, they can be routed around. A crucial figure of merit becomes the site percolation threshold for the underlying lattice (where we consider the overall lattice to be comprised of a convolution between the underlying lattice and the subsystem) -- if the probability of an error in a given block of spins is below the percolation threshold, then long range communication is certainly possible, although it does not guarantee communication between any given input and output node is possible. We can even choose the underlying geometry to optimise this tolerance using, for instance, a triangular lattice in 2D.

\begin{figure}[!tb]
\begin{center}
\includegraphics[width=0.45\textwidth]{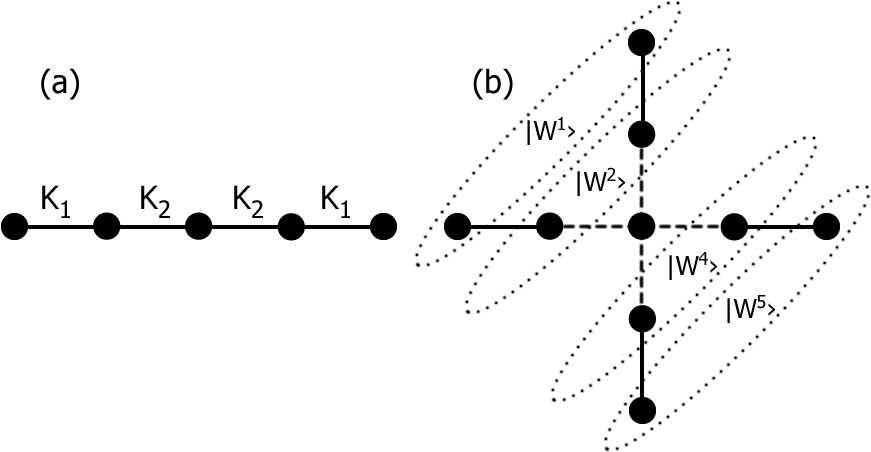}
\end{center}\vspace{-0.5cm}
\caption{(a) A perfect state transfer coupling scheme for a chain of 5 qubits. (b) Conversion of (a) into a star topology ($d=2$), with $\ket{n}$ in (a) transforming into $\ket{W^n_0}$. Dashed coupling strengths are scaled to $K_2/\sqrt{d}$ for $2d$ branches.} \label{fig:star}\vspace{-0.5cm}
\end{figure}

{\em Dynamics within a Subsystem:} We will now give a simple construction of a subsystem structure based on the design of perfect state transfer chains. We start from the solution to perfect state transfer for a chain of $M\geq 5$ qubits ($M$ odd). Such a scheme can be written as
$$
H_{\text{chain}}=\sum_{n=1}^{M-1}K_n(\ket{n}\bra{n+1}+\ket{n+1}\bra{n}),
$$
and exhibits perfect transfer in time $t_0$, i.e. $e^{-iH_{\text{chain}}t_0}\ket{n}=\ket{M+1-n}$. Using the techniques introduced in \cite{Kay:2005b,kay-2006b}, this can be redesigned into a star topology of $2d$ branches, where the central coupling becomes $J_{(M-1)/2}/\sqrt{d}$ (see Fig.~\ref{fig:star}). Under this transformation, the single excitation states of the chain, $\ket{n}$, map to $\ket{W_0^n}$ (for $n\neq (M+1)/2$) where
$$
\ket{W_k^n}=\frac{1}{\sqrt{d}}\sum_{j=a}^de^{\frac{2\pi ijk}{d}}\ket{nj}
$$
and each of the split chains is indexed from $a$ to $d$, i.e.~the extremal nodes are $1a$ to $1d$ and $Ma$ to $Md$ (although, due to symmetry, it is entirely irrelevant which are labelled as which). Thus, by definition,
$$
e^{-iH_{\text{star}}t_0}\ket{W^n_0}=\ket{W^{M+1-n}_0}.
$$
Now we need to know the dynamics of the other states $\ket{W^n_k}$ in the same time. The Hamiltonian $H_{\text{star}}$ decomposes into a further direct sum structure of fixed $k$, each with a hopping Hamiltonian
$$
\sum_{n=1}^{(M-3)/2}K_n(\ket{W^n_k}\bra{W^{n+1}_k}+\ket{W^{n+1}_k}\bra{W^n_k}).
$$
This is exactly the same effective Hamiltonian as for $H_{\text{chain}}$ acting on states of the form $(\ket{m}-\ket{M+1-m})/\sqrt{2}$. Consequently, for $k\neq 0$, 
$$
e^{-iH_{\text{star}}t_0}\ket{W_k^n}=-\ket{W_k^n}.
$$
By restricting to these intervals of $t_0$, we find a very simple description of the unitary evolution of states on the star subsystem, which we can make use of for designing the routing protocols within the subsystem.

{\em Routing within a Subsystem:} Our aim is now to show how to route an input state within a subsystem, i.e.~to transmit it from spin $1j$ to $1l$. The input and output states can be written as $\ket{1j}=\frac{1}{\sqrt{d}}\sum_ke^{-\frac{2\pi ijk}{d}}\ket{W^1_k}$, which differ only by the relative phases of the $\ket{W^1_k}$ states. We can start doing this by evolving for time $t_0$, which creates
$$
\frac{1}{\sqrt{d}}\left(\ket{W^M_0}-\sum_{k\neq 0}e^{-\frac{2\pi ijk}{d}}\ket{W^1_k}\right).
$$
By applying a phase gate of phase $\theta$, $Z^{(\theta)}$, on each of the spins $1j$, it is converted to
$$
\frac{1}{\sqrt{d}}\left(\ket{W^M_0}-e^{i\theta}\sum_{k\neq 0}e^{-\frac{2\pi ijk}{d}}\ket{W^1_k}\right).
$$
After another transfer time $t_0$, this yields
$$
\frac{1}{\sqrt{d}}\left(\sum_{k\neq 0}e^{\frac{-2\pi ijk}{d}}\ket{W^1_k}+e^{-i\theta}\ket{W^1_0}\right),
$$
so this shifts the relative phase of the $\ket{W_0^1}$ component. Now all we have to do is apply local phase gates on each spin $1j$ with the cumulative effect that
$$
\left(\prod_{j=a}^dZ^{(2\pi j/d)}_j\right)\ket{W_k}=\ket{W_{k+1\text{ mod }d}},
$$
enabling us to permute through each $\ket{W_k}$ and alter its phase, creating the state we need after only time $2dt_0$. Hence, we can route from any node $1j$ to any other $1l$.

In fact, we can observe that reduced control suffices to transfer a state. Consider the ability to apply $Z$ gates on the input and output spins only. There is an effective Hamiltonian term coming from the Hamiltonian evolution for the perfect state transfer time,
$$
H_1=\sum_n(\ket{W^{M+1-n}_0}+\ket{W^n_0})(\bra{W^{M+1-n}_0}+\bra{W^n_0})-\identity,
$$
whereas the two $Z$ fields are written as $H_2=\identity-2\proj{1j}$ and $H_3=\identity-2\proj{Ml}$. Now observe that
$$
\left[H_3,\left[H_1,H_2\right]\right]=\frac{4}{d^2}\left(\ket{1j}\bra{Ml}+\ket{Ml}\bra{1j}\right),
$$
proving that the transfer can be achieved between any pair of controlled nodes, without control over the others.

With local control over magnetic fields on each output node of the subsystem, we can show full control over the single excitation subspace across spins $1a$ to $1d$ and $Ma$ to $Md$. This allows us, for instance, to create entangled states, and this is particularly simple if $d$ is even because then we can use the phase changing protocol to create
$$
\frac{1}{\sqrt{d}}\sum_je^{i\pi(-1)^j/4}\ket{W_j}=\frac{1}{\sqrt{2}}(\ket{1d}+i\ket{1\smallfrac[d]{2}}).
$$

While there are infinitely many solutions to the perfect state transfer problem using a fixed Hamiltonian \cite{transfer_comment,kay:review}, and our construction is general enough to take advantage of any of them, there is a particularly beautiful choice if $d=3$ (i.e.~2D triangular or cubic lattice). Here we select the standard perfect state transfer solution for $M=5$ \cite{Christandl}, i.e.~$K_1=\sqrt{2}$ and $K_2=\sqrt{3}$, meaning that the regular network that we construct has every coupling strength taking on the same modulus. Furthermore, this solution is the most efficient solution to the state transfer problem with regard to a number of parameters \cite{Kay:2005e,yung:06,kay:review}.

{\em Conclusions:} In this paper, we have shown how regular networks which are local in any number of spatial dimensions, $d$, can be designed to route quantum states from arbitrary nodes in a time that is linear in the distance to be covered. Moreover, multiple states can be transferred at once, meaning that, even in the 1D case, a transfer rate can be realised which is in excess of that achievable in perfect state transfer schemes \cite{Kay:2005e,kay:review}. To achieve this required a level of control that was never explicitly utilised in previous perfect state transfer schemes until its recent observation in \cite{Kay:10}, but was implicitly present for the addition and removal of states. Moreover, the fields that we utilise can be applied in a global way, addressing, for instance, every second or third spin at regular intervals. These properties therefore bring such a scheme much closer to reality.

There are still a number of features that have been found in the study of perfect state transfer that one might like to recover in the network routing scenario, such as independence of initial state \cite{kay-2006b,paternostro:1}, or no requirement for `fast' pulses \cite{haselgrove04,Kay:08b}. It would be interesting to design a network that allows us to entangle the individually generated pairs to create long-range entanglement in a single time step (much as you would in a cluster state). There are also questions to be addressed regarding cooperating parties -- it is currently assumed that all parties cooperate, but how many error prone participants can be tolerated? A starting point would be to quantify the ranks of the controllable subspaces of the subsystem structure as more controlling magnetic fields are added. If the rank increases from $n$ magnetic fields to $n+1$, then that reveals that if $n$ parties are attempting to cooperate to send a state, then one of the passive parties can certainly ruin the transfer by moving the state into a subspace that is not controlled by the other parties. Can this be protected against through the use of quantum or classical Byzantine agreement protocols, or perhaps just through the use of error correcting codes? If uncooperative parties can be identified, then they can be treated in much the same was as faults in the manufacture of the network, and routed around.  Finally, we hope that this particularly elegant construction can find application in other contexts.

PJP acknowledges funding from an EPSRC PhD studentship. ASK is supported by Clare College, Cambridge.

\end{document}